\documentclass{webofc}
\usepackage[varg]{txfonts}   % Web of Conferences font
\usepackage{mathrsfs}
\begin{document}
\title{Gravitational Wave Polarizations in $f(R)$ Gravity and Scalar-Tensor Theory}

\author{\firstname{Yungui} \lastname{Gong}\inst{1}\fnsep\thanks{\email{yggong@hust.edu.cn}, invited talk given by this author} \and
        \firstname{Shaoqi} \lastname{Hou}\inst{1}\fnsep\thanks{\email{shou1397@hust.edu.cn}}
        % etc.
}

\institute{School of Physics, Huazhong University of Science and Technology,
Wuhan, Hubei 430074, China}

\abstract{
  The detection of gravitational waves by the Laser Interferometer Gravitational-Wave Observatory opens a new era to use gravitational waves to test alternative theories of gravity. We investigate the polarizations of gravitational waves in $f(R)$ gravity and Horndeski theory, both containing scalar modes. These theories predict that in addition to the familiar $+$ and $\times$ polarizations, there are transverse breathing and longitudinal polarizations excited by the massive scalar mode and the new polarization is a single mixed state.
  It would be very difficult to detect the longitudinal polarization by interferometers, while pulsar timing array may be the better tool to detect the longitudinal polarization.
}
\maketitle
\section{Introduction}
\label{intro}

In September 14th, 2015, the Laser Interferometer Gravitational-Wave Observatory (LIGO) Scientific Collaboration and Virgo Collaboration announced the first detection of gravitational waves (GWs), namely GW150914 \cite{gw150914}, nearly 100 years after Einstein's prediction of GWs based on his theory of general relativity (GR). Soon after, two more GW events, GW151226 \cite{gw151226}, later that year, and  GW170104 \cite{gw170104} early this year were observed. A new era is thus already open, and GW is a tool for testing GR and alternative theories of gravity, for GWs in alterative theories of gravity are predicted to have different polarization contents. The polarizations of null GWS with the propagation speed $c$ were
classified by the Newman-Penrose variables \cite{Newman1962} in \cite{Eardley1973}. However, the application of the
classification by the Newman-Penrose variables to
the polarizations of GWs with the propagation speed different from $c$
leads to confusing results \cite{Liang:2017gwa}.
Although LIGO presently cannot tell the polarizations of the observed GWs, it will become possible in the future when more interferometers are in operation.
Other detectors, such as pulsar timing arrays, are also capable of distinguishing the polarization content of GWs \cite{Hellings1983}.

Since the birth of GR, various alterative theories of gravity have been proposed for different motivations. One of the motivations is
to cure the nonrenormalizability of GR. For instance, the inclusion of quadratic terms in Riemann tensor $R_{\mu\nu\rho\sigma}$ in the action makes gravity renormalizable \cite{Utiyama1962,Stelle1977}, and one simple realization of this idea is $f(R)$ gravity \cite{Buchdahl1970}.
Another motivation is to explain the present accelerating expansion of the universe \cite{Adam1998,Perlmutter1999}. The addition of new gravitational degrees of freedom might do the work, and the scalar-tensor theory is the simplest alternative metric theory of gravity which contains a scalar field $\phi$ besides the metric tensor $g_{\mu\nu}$ to describe gravity. In 1974, Horndeski constructed the most general scalar-tensor theory of gravity whose action has higher derivatives of $\phi$ and $g_{\mu\nu}$, but yields at most second order differential equations of motion \cite{Horndeski1974}. Therefore, there is no Ostrogradsky instability \cite{Ostrogradsky1850}.

In this talk, we will discuss the polarizations of GWs predicted by $f(R)$ gravity \cite{Liang:2017gwa} and Horndeski theory \cite{Hou:2017gwb}. In Section \ref{sec-fr}, we focus on $f(R)$ gravity. We will first resolve the debate on how many physical degrees of freedom are contained in $f(R)$ gravity, recently raised by Refs. \cite{Corda2007,Corda2008,Kausar2016,Myung2016}. Next, we determine the polarizations of $f(R)$ gravity using geodesic deviation equations. In Section \ref{sec-st}, we consider the polarizations of GWs in the scalar-tensor theory, in particular, Horndeski theory. We first determine the polarizations with similar method used for $f(R)$ gravity, then we consider the possibility to detect the polarizations using the pulsar timing arrays and the interferometers. Finally, there is a brief conclusion.

\section{Gravitational Wave Polarizations in $f(R)$ Gravity}
\label{sec-fr}

Utiyama and DeWitt showed that the addition of the quadratic terms $R_{\mu\nu\rho\sigma}R^{\mu\nu\rho\sigma}$ and $R^2$ to the Einstein-Hilbert action makes gravity renormalizable at one-loop level \cite{Utiyama1962}, and then, Stelle proved the renormalizability at all loop levels \cite{Stelle1977}. In four dimensions, the volume integration of the Gauss-Bonnet term $\mathcal{G}=R_{\mu\nu\rho\sigma}R^{\mu\nu\rho\sigma}-4R_{\mu\nu}R^{\mu\nu}+R^2$ vanishes for a simply connected spacetime, as usually considered.
So the general quadratic terms can be given by
\begin{equation}\label{quagen}
  \frac{1}{2\kappa}\int d^4x\sqrt{-g}(\alpha'R_{\mu\nu}R^{\mu\nu}+\alpha R^2),
\end{equation}
where $\kappa=8\pi G_\mathrm{N}$ with $G_\mathrm{N}$ Newton's constant, $g$ is the determinant of $g_{\mu\nu}$ and $\alpha',\alpha$ are constants. Setting $\alpha'=0$, one obtains a model which was first proposed by Starobinsky as an inflationary model \cite{Starobinsky1980}, and is consistent with the observations of \textit{Planck} \cite{Planck2016}. One may thus generalize this action by considering a generic function $f$ of $R$ \cite{Buchdahl1970},
\begin{equation}\label{fract}
  S=\frac{1}{2\kappa}\int d^4x\sqrt{-g}f(R).
\end{equation}
A special model with $f(R)=R+\beta R^{-1}$ was applied to explain the late time cosmic acceleration \cite{Vollick2003,Nojiri2003,Carroll2004,Flanagan2004}, but the solar system tests have ruled it out \cite{Chiba2013,Erickcek2006}. So more viable $f(R)$ models were proposed recently \cite{Hu2007,Starobinsky2007,Cognola2008,Nojiri2008,Capozziello2009,Myrzakulov2015,Yi2016}.

In fact, $f(R)$ gravity is equivalent to a scalar-tensor theory \cite{Hanlon1972,Teyssandier1983}, as the action can be rewritten as
\begin{equation}\label{frst}
  S=\frac{1}{2\kappa}\int d^4x\sqrt{-g}[f(\varphi)+(R-\varphi)f'(\varphi)],
\end{equation}
where $f'(\varphi)=d f(\varphi)/d\varphi$. It can be easily shown that $\varphi=R$ on-shell. The polarization content of GWs in $f(R)$ gravity
and the detection have been studied in Refs. \cite{Corda2007,Corda2008,Capozziello2008,Capozziello2010}.
Authors of Refs. \cite{Corda2007,Corda2008} found out that the massive scalar mode induces the longitudinal polarization, and they claimed that there are four degrees of freedom using the Newman-Penrose (NP) formalism \cite{Newman1962,Eardley1973}. Kausar \textit{et. al.} supported this claim by arguing that the traceless condition cannot be implemented \cite{Kausar2016}. However, Myung's work shows that there is no issue with implementing the transverse traceless condition \cite{Corda2008}, and there are only three degrees of freedom in $f(R)$ gravity \cite{Myung2016}.

In Ref.~\cite{Liang:2017gwa}, we investigated the polarizations of GWs in $f(R)$ gravity
and attempted to resolve the debate on how many degrees of freedom propagating in this theory.
As it will become clear soon, there are three physical propagating degrees of freedom in $f(R)$ gravity.
Therefore, there are the familiar $+$ and $\times$ polarizations as in GR, and the transverse
and longitudinal polarizations excited by the massive scalar mode $\varphi=R$.
We also pointed out that the original NP formalism devised in Ref.~\cite{Eardley1973} for
identifying the polarizations of GWs in a generic metric theory of gravity cannot be
simply applied to massive GWs. In order to reveal the polarizations, one simply calculates
the geodesic deviations caused by the GW, provided that test particles follow geodesics, as usually assumed.

\subsection{Equations of Motion}
\label{sec-freom}

The field equations can be obtained by the variational principle,
\begin{equation}
\label{freinseq1}
f'(R)R_{\mu\nu}-\frac{1}{2}f(R)g_{\mu\nu}-\nabla_\mu\nabla_\nu f'(R)+g_{\mu\nu}\Box f'(R)=0,
\end{equation}
where $\Box=g^{\mu\nu}\nabla_\mu\nabla_\nu$. Taking the trace of Eq. \eqref{freinseq1}, we get
\begin{equation}
\label{freinseq2}
f'(R)R+3\Box f'(R)-2f(R)=0.
\end{equation}
For the particular model $f(R)=R+\alpha R^2$, Eq. \eqref{freinseq1} becomes
\begin{equation}
\label{frperteq3}
R_{\mu\nu}-\frac{1}{2}\eta_{\mu\nu}R-2\alpha\left(\partial_\mu\partial_\nu R-\eta_{\mu\nu}\Box R\right)=0,
\end{equation}
Take the trace of Eq. \eqref{frperteq3} or using Eq. \eqref{freinseq2}, we have
\begin{equation}
\label{frperteq4}
(\Box-m^2)R=0,
\end{equation}
where $m^2=1/(6\alpha)$ with $\alpha>0$. The graviton mass $m$ has been bounded from above by GW170104 as
$m<m_b=7.7\times10^{-23}\text{ eV}/c^2$ \cite{gw170104}, and the observation of the dynamics of the galaxy cluster puts a more stringent limit, $m<2\times10^{-29}\text{ eV}/c^2$ \cite{Goldhaber1974}.

To obtain the GW solutions in the flat spacetime background, perturb the metric around the Minkowski metric $g_{\mu\nu}=\eta_{\mu\nu}+h_{\mu\nu}$ to the first order of $h_{\mu\nu}$, and introduce an auxiliary metric tensor
\begin{equation}
\label{canhtt}
\bar{h}_{\mu\nu}=h_{\mu\nu}-\frac{1}{2}\eta_{\mu\nu}h-2\alpha\eta_{\mu\nu} R,
\end{equation}
which  transforms in an infinitesimal coordinate transformation $x^\mu\rightarrow x'^\mu=x^\mu+\epsilon^\mu$ in the following way,
\begin{equation}
\label{coortranfeq23}
\bar{h}_{\mu\nu}'=\bar{h}_{\mu\nu}-\partial_\mu\epsilon_\nu-\partial_\nu\epsilon_\mu+\eta_{\mu\nu}\partial_\rho\epsilon^\rho,
\end{equation}
where to the first order of perturbation, the index was raised or lowered by the Minkowski metric $\eta_{\mu\nu}$, i.e., $\epsilon_\mu=\eta_{\mu\nu}\epsilon^\nu$.
If we choose $\epsilon_\mu$ so that it satisfies $\Box\epsilon_\nu=\partial^\mu\bar{h}_{\mu\nu}$ with $\Box=\partial^\mu\partial_\mu$ from now on, then we get the Lorenz gauge condition $\partial^\mu \bar{h}_{\mu\nu}'=0$. Note that there is still some residual gauge freedom, i.e., $x^{\mu\prime}=x^\mu+\xi^\mu$ with $\Box\xi^\mu=0$. If $\xi^\mu$ also satisfies the relation $\partial_\mu\xi^\mu=-\bar{h}/2$, then  $\bar{h}'=0$ holds. Therefore, it is always possible to choose the transverse traceless gauge condition
\begin{equation}
\label{gaugeeq1}
\partial^\mu \bar{h}_{\mu\nu}=0,\quad \bar{h}=\eta^{\mu\nu}\bar{h}_{\mu\nu}=0.
\end{equation}
With this gauge condition,  some algebraic manipulations lead to
\begin{equation}
\label{frperteq8}
\Box\bar{h}_{\mu\nu}=0.
\end{equation}
Therefore, the equations of motion are Eqs. (\ref{frperteq4}) and (\ref{frperteq8}).

The plane wave solution can be obtained immediately,
\begin{gather}
  \bar h_{\mu\nu}=e_{\mu\nu}\exp(iq_\mu x^\mu)+c.c.,\label{sols-h}\\
  R=\phi_1\exp(ip_\mu x^\mu)+c.c.,\label{sols-r}
\end{gather}
where $c.c.$ stands for the complex conjugation, $e_{\mu\nu}$ and $\phi_1$ are the amplitudes with $q^\nu e_{\mu\nu}=0$ and $\eta^{\mu\nu}e_{\mu\nu}=0$, and $q_\mu$ and $p_\mu$ are the wave numbers satisfying
\begin{equation}\label{wvnm}
  \eta^{\mu\nu}q_\mu q_\nu=0, \quad
  \eta^{\mu\nu}p_\mu p_\nu=-m^2.
\end{equation}

\subsection{Physical Degrees of Freedom}

In this subsection, we will find the number of physical degrees of freedom in $f(R)$ gravity,
using two different methods: examining the energy current carried by GWs and carrying out the Hamiltonian analysis.

\subsubsection{Energy Current of GWs}

The first method to determine the number of physical degrees of freedom is to calculate the energy current carried by the GW propagating, for example, in the $+z$ direction. Let us consider the null GWs with $q^\mu=\omega(1,0,0,1)$, then the energy current is given by \cite{Berry:2011pb},
\begin{equation}\label{ecs}
\begin{split}
  t_{0z}=&\frac{1}{\kappa}\langle G_{0z}^{(1)}-G_{0z}\rangle\\
  =&\frac{1}{2\kappa}\left\langle \omega^2\left[\left(\frac{e_{xx}-e_{yy}}{2}\right)^2+e_{xy}^2\right]+48\alpha^2R_{,0}R_{,z}\right\rangle,
  \end{split}
\end{equation}
where $G_{\mu\nu}^{(1)}$ is the first order Einstein tensor. Note that to obtain the above result,
the solution (\ref{sols-h}) is used, but the traceless condition is dropped. Eq. (\ref{ecs}) makes it clear that a
null GW for which $e_{xx}-e_{yy}$ and $e_{xy}$ are both zero does not transport energy. So if a null wave has a
nonvanishing trace $\bar h$ such that $e_{xx}+e_{yy}\ne0$, it does not carry energy, which implies
that the trace $\bar h$ is not a dynamical degree of freedom. The null GW $\bar h_{\mu\nu}$ is
physically transverse and traceless, so has two degrees of freedom as in GR. In addition,
the Ricci scalar $R$ is the third degree of freedom. So totally, there are three degrees of freedom.

\subsubsection{Hamiltonian Analysis}

The Hamiltonian analysis of $f(R)$ gravity has been done in Refs. \cite{Ezawa1999,Ezawa2006,Deruelle2009,Deruelle2010,Sendouda2011,Olmo2011,Ohkuwa2015}. In our work, we did the Hamiltonian analysis with the action (\ref{frst}) for simplicity. With the Arnowitt-Deser-Misner (ADM) foliation \cite{Arnowitt1962,arnowitt_republication_2008}, the metric takes the standard form
\begin{equation}\label{lem}
  ds^2=-N^2dt^2+h_{jk}(dx^j+N^jdt)(dx^k+N^kdt),
\end{equation}
where $N,N^j,h_{jk}$ are the lapse function, the shift function and the induced metric on the constant $t$ slice $\Sigma_t$, respectively. Let $n_\mu=-N\nabla_\mu t$ be the unit normal to $\Sigma_t$, and the exterior curvature is $K_{\mu\nu}=\nabla_\mu n_\nu+n_\mu n^\rho\nabla_\rho n_\nu$. In terms of ADM variables and setting $\kappa=1$, the action (\ref{frst}) is
\begin{equation}
  S = \int d^4xN\sqrt{h}\Big[\frac{1}{2}f'(\mathscr R-\varphi)+\frac{1}{2}f+\frac{1}{2}f'(K_{jl}K^{jl}-K^2)+\frac{K}{N}(N_jD^jf'-f''\dot\varphi)+D_j f' D^j\ln N\Big],
\end{equation}
where  $\mathscr R$ is the Ricci scalar for $h_{jk}$ and $K=h^{jk}K_{jk}$. In this action, there are 11 dynamical variables: $N,N_j,h_{jk}$ and $\varphi$. Four primary constraints are immediately recognized, i.e., the conjugate momenta for $N$ and $N_j$ vanish weakly,
\begin{equation}\label{prcon}
    \pi^N = \frac{\delta S}{\delta \dot N}\approx0,\quad  \pi^j=\frac{\delta S}{\delta \dot N_j}\approx0.
\end{equation}
The conjugate momenta for $h_{jk}$ and $\varphi$ can also be obtained, and the Legendre transformation results in the following Hamiltonian,
\begin{equation}
  H=\int_{\Sigma_t}d^3x\sqrt{h}(NC+N_jC^j),
\end{equation}
where we dropped the boundary terms. Thus, the consistence conditions yield four secondary constraints, i.e., $C\approx0$ and $C^j\approx0$ \footnote{For details, please refer to Ref.~\cite{Liang:2017gwa}}, and it can be shown that there are no further secondary constraints. It can also be checked that all the constraints are of the first class, so the number of physical degrees of freedom of $f(R)$ gravity is
\begin{equation}\label{dof}
  n=\frac{22-8\times2}{2}=3,
\end{equation}
as expected.

\subsection{Polarization Content}

To reveal the polarization content of GWs in $f(R)$ gravity, let us calculate the geodesic deviation equations caused by the GW propagating in the $+z$ direction with the wave vectors given by
\begin{equation}\label{wvecs}
  q^\mu=\omega(1,0,0,1),\quad p^\mu=(\Omega,0,0,\sqrt{\Omega^2-m^2}).
\end{equation}
Inverting Eq. (\ref{canhtt}), one obtains the metric perturbation,
\begin{equation}\label{metper}
  h_{\mu\nu}=\bar h_{\mu\nu}(t-z)-2\alpha\eta_{\mu\nu}R(vt-z),
\end{equation}
where $v=\sqrt{\Omega^2-m^2}/\Omega$. It is expected that $\bar h_{\mu\nu}$ induces the $+$ and $\times$ polarizations. So let us investigate the polarization state caused by the massive scalar field by setting $\bar h_{\mu\nu}=0$. The geodesic deviation equations are
\begin{equation}\label{geodev}
  \ddot{x}=\alpha\ddot R x,\quad \ddot y=\alpha\ddot{R}y,\quad \ddot z=-\alpha m^2R z=-\frac{1}{6}R z.
\end{equation}
Therefore, the massive scalar field induces a mix of the pure longitudinal and the breathing modes.

The NP formalism \cite{Eardley1973} cannot be applied to infer the polarization content
of $f(R)$ gravity because the NP formalism was formulated for null GWs.
In fact, the calculation shows that $\Psi_2$ is zero. According to the NP formalism, $\Psi_2=0$ means the absence of the longitudinal polarization.
From Eq. \eqref{geodev}, we see the existence of the longitudinal polarization.
Since the six polarization states are completely determined by the electric part of the Riemann tensor $R_{itjt}$, we can still use the six
polarizations classified by the NP formalism as the base states. In terms of these polarization base states, the polarization state caused by the massive
scalar field is a mix of the longitudinal and the breathing modes.
Since there is no massless limit in $f(R)$ gravity, so we consider more general massive scalar-tensor theory of gravity.

\section{Gravitational Wave Polarizations in Scalar-Tensor Theory}\label{sec-st}

As stated before, $f(R)$ gravity is equivalent to a scalar-tensor gravity. We extended our work \cite{Liang:2017gwa} to the scalar-tensor theory,
and study the polarization content of GWs in Horndeski theory \cite{Hou:2017gwb}. The action is given by\cite{Horndeski1974},
\begin{equation}
\label{acth}
  S=\int d^4x\sqrt{-g}(L_2+L_3+L_4+L_5),
\end{equation}
where
\begin{gather*}
L_2=K(\phi,X),\quad L_3=-G_3(\phi,X)\Box \phi, \quad L_4=G_4(\phi,X)R+G_{4,X}\left[(\Box\phi)^2-(\nabla_\mu\nabla_\nu\phi)(\nabla^\mu\nabla^\nu\phi)\right], \nonumber\\
L_5=G_5(\phi,X)G_{\mu\nu}\nabla^\mu\nabla^\nu\phi-\frac{1}{6}G_{5,X}\left[(\Box\phi)^3-3(\Box\phi)(\nabla_\mu\nabla_\nu\phi)(\nabla^\mu\nabla^\nu\phi)+2(\nabla^\mu\nabla_\alpha\phi)(\nabla^\alpha\nabla_\beta\phi)(\nabla^\beta\nabla_\mu\phi)\right].
\end{gather*}
Here, $X=-\nabla_\mu\phi\nabla^\mu\phi/2$, $\Box\phi=\nabla_\mu\nabla^\mu\phi$,
the functions $K$, $G_3$, $G_4$ and $G_5$ are arbitrary functions of $\phi$ and $X$, and $G_{j,X}(\phi,X)=\partial G_j(\phi,X)/\partial X$ with $j=4,5$. Horndeski theory includes several interesting theories as its subclasses.
For example, to reproduce $f(R)$ gravity, one can set $G_3=G_5=0$, $K=f(\phi)-\phi f'(\phi)$ and $G_4=f'(\phi)$ with $f'(\phi)=d f(\phi)/d\phi$.

\subsection{Linearized Equations of Motion}

The equations of motion can be derived with the variational principle. Interested readers should be referred to Ref.~\cite{Kobayashi2011ginf}. We are interested in the GW solutions in the flat spacetime background for which  $g_{\mu\nu}=\eta_{\mu\nu}$ and $\phi=\phi_0$ with a constant $\phi_0$. This requires that $K(\phi_0,0)=0$ and $K_{,\phi_0}=\partial K(\phi,X)/\partial\phi|_{\phi=\phi_0,X=0}=0$. Now the fields are expanded around the background, $g_{\mu\nu}=\eta_{\mu\nu}+h_{\mu\nu}$ and $\phi=\phi_0+\varphi$. To the first order of approximation, the equations of motion are
\begin{equation}\label{eqht}
(\Box-m^2)\varphi = 0,\quad
G_{\mu\nu}^{(1)}-\frac{G_{4,\phi_0}}{G_4(0)}(\partial_\mu\partial_\nu\varphi-\eta_{\mu\nu}\Box\varphi)=0,
\end{equation}
where $G_4(0)=G_4(\phi_0,0)$, $K_{,X_0}=\partial K(\phi,X)/\partial X|_{\phi=\phi_0,X=0}$ and the mass squared of the scalar field is
\begin{equation}
\label{msq}
m^2=-\frac{K_{,\phi_0\phi_0}}{K_{,X_0}-2G_{3,\phi_0}+3G_{4,\phi_0}^2/G_{4}(0)}.
\end{equation}

Similar to Eq. (\ref{canhtt}), define a field $\tilde h_{\mu\nu}$,
\begin{equation}
\label{auht}
\tilde h_{\mu\nu}=h_{\mu\nu}-\frac{1}{2}\eta_{\mu\nu}\eta^{\alpha\beta}h_{\alpha\beta}-\frac{G_{4,\phi_0}}{G_{4}(0)}\eta_{\mu\nu}\varphi,
\end{equation}
and choose the transverse traceless gauge $\partial_\mu \tilde h^{\mu\nu}=0$, $\eta^{\mu\nu}\tilde h_{\mu\nu}=0$ by
using the freedom of coordinate transformation, then the linearized equations \eqref{eqht} become two wave equations,
\begin{gather}
\label{eq-sceqf}
(\Box-m^2)\varphi = 0,\\
\label{eq-eineqf}
\Box\tilde h_{\mu\nu} = 0.
\end{gather}

\subsection{Polarization Content}

A inspection of Eq. (\ref{eq-eineqf}) makes it clear that the field $\tilde{h}_{\mu\nu}$ denotes the usual massless gravitons and it has two polarization states, the $+$ and $\times$ modes.
The plane waves (\ref{sols-h}) and (\ref{sols-r}) are also the solutions to Eqs. (\ref{eq-eineqf}) and (\ref{eq-sceqf}).
For GWs propagating in the $+z$ direction with waves vectors given by Eq. (\ref{wvecs}), one gets the following nonvanishing NP variables
\begin{gather}\label{nps}
 \Psi_4=-\omega^2(\tilde h_{xx}-i\tilde h_{xy}),\quad
\Phi_{22}=\frac{(\Omega+\sqrt{\Omega^2-m^2})^2}{4}\sigma\varphi,
  \\
\Phi_{00}=\frac{4(\Omega-\sqrt{\Omega^2-m^2})^2}{(\Omega+\sqrt{\Omega^2-m^2})^2}\Phi_{22},\quad
\Phi_{11}=-\Lambda=\frac{4m^2}{(\Omega+\sqrt{\Omega^2-m^2})^2}\Phi_{22},
\end{gather}
where $\sigma=G_{4,\phi_0}/G_4(0)$. This result shows that the massive scalar field makes
$\Phi_{22}$, $\Phi_{00}$, $\Phi_{11}$ and $\Lambda$ nonzero and they are all proportional to $\varphi$.
However, for the NP variables associated with the Weyl tensor, only $\Psi_4\ne0$ and in particular, $\Psi_2=0$.
In fact, $\Psi_4$ denotes the usual $+$ and $\times$ modes for the massless gravitons. If the scalar field is massless,
only $\Phi_{22}$ is nonvanishing and it represents the transverse breathing mode \cite{Eardley1973}.

Next, let us focus on the GW induced by the scalar field $\varphi$.
Setting $\tilde h_{\mu\nu}=0$ and inverting Eq. (\ref{auht}) to obtain $h_{\mu\nu}$ in terms of $\varphi$, the geodesic deviation equations are
\begin{equation}\label{geodev2}
  \ddot x=-\frac{1}{2}\Omega^2\sigma\varphi x, \quad
  \ddot y=-\frac{1}{2}\Omega^2\sigma\varphi y, \quad
  \ddot z=-\frac{1}{2}m^2\sigma\varphi z.
\end{equation}
These expressions take essentially the same form as Eq. (\ref{geodev}). Therefore, the massive scalar field
excites the longitudinal and transverse breathing polarizations, while the massless scalar field excites only the transverse breathing polarization.
As discussed in the previous section, $\Psi_2=0$ means the absence of the longitudinal polarization, however, there exists
the longitudinal mode even though $\Psi_2=0$, so the NP formalism derived for null GWs cannot be directly applied
to the massive GWs, and again
the polarization state caused by the massive
scalar field is a mix of the longitudinal and the breathing modes.

\subsection{Experimental Tests}

\begin{figure}[htp]
  \centering
  \includegraphics[width=0.4\textwidth,clip]{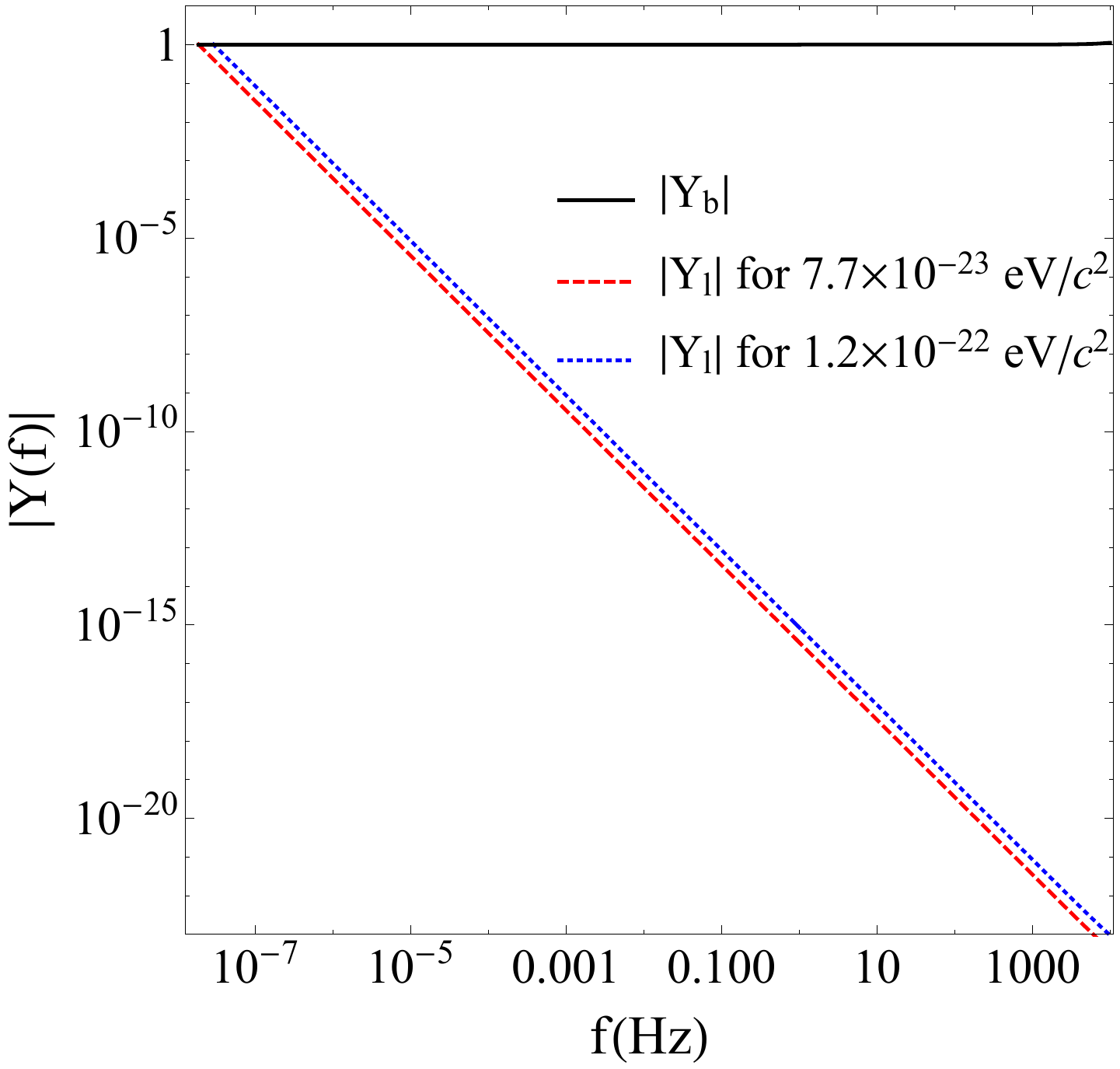}
  \caption{The absolute values of the longitudinal and transverse response functions $|Y_l(f)|$ and $|Y_b(f)|$
  as functions of $f$ for aLIGO  to a scalar gravitational wave with the masses $1.2\times10^{-22}\,\text{eV}/c^2$ \cite{gw150914} (dotted blue curve) and $7.7\times 10^{-23}$ eV/$c^2$ \cite{gw170104} (dashed red curve).  The solid curve denotes $|Y_b(f)|$.}\label{fig-yl}
\end{figure}

In the interferometer, a photon is emanated from the beam splitter, bounced back by the mirror and returns to the beam splitter.
The propagation time when there is no GW does not equal the one when there is.
Following \cite{Rakhmanov2005,Corda2007}, we calculate the interferometer response functions of the transverse and longitudinal polarizations in the frequency domain.
To calculate the response functions, we assume the beam splitter is placed at the origin of the coordinate system.
Figure \ref{fig-yl} shows the absolute value of the longitudinal and transverse response functions for
aLIGO  to a scalar GW with the masses $1.2\times10^{-22}\,\text{eV}/c^2$ \cite{gw150914} and $7.7\times 10^{-23}$ eV/$c^2$ \cite{gw170104}. From this graph, it is clear that the response functions for the longitudinal polarization are smaller than that of the transverse breathing modes by several orders of magnitude at high frequency regime, so it is difficult to test the existence of longitudinal polarization by interferometer such as aLIGO.

\begin{figure}
  \centering
  \includegraphics[width=0.4\textwidth]{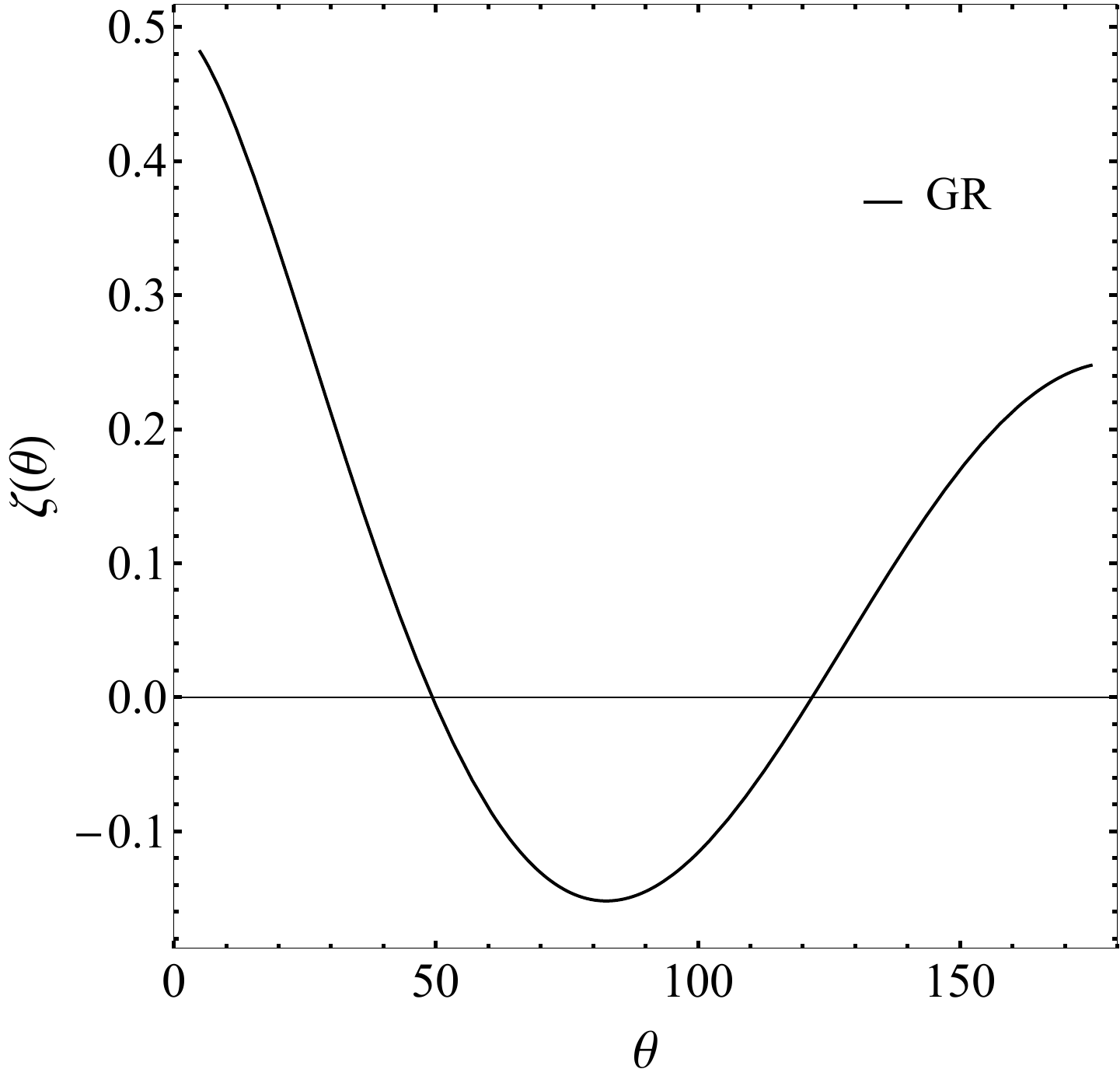}
  \includegraphics[width=0.4\textwidth]{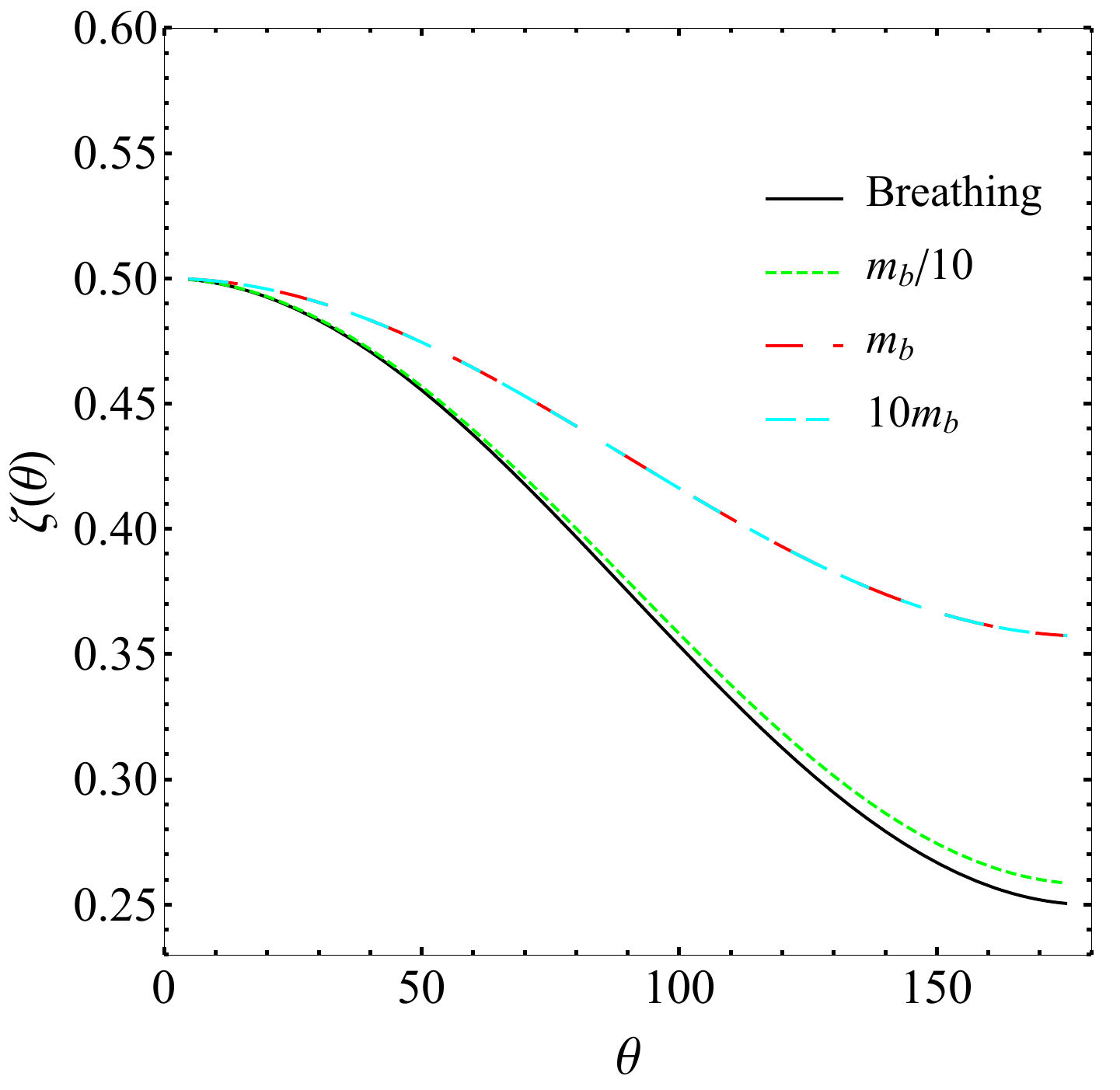}
  \caption{The normalized correlation function $\zeta(\theta)=C(\theta)/C(0)$ as a function of $\theta$ for the massless field $\tilde h_{\mu\nu}$ (the left panel) and the scalar field  $\varphi$ (the right panel).}\label{fig-cor}
\end{figure}

It is possible to tell what the polarization content of GWs is by analyzing the data of the pulsar timing arrays \cite{Hellings1983,Lee2008ptac,Lee2010pta,Chamberlin2012,Lee2013,Yunes2013lrr,Gair2014,Gair2015}.
The stochastic gravitational wave background causes the pulse time-of-arrival (TOA) residuals $\tilde{R}(t)$ of pulsars which can be measured \cite{Hellings1983}.
%The plane wave solution (\ref{sols-h}) and (\ref{sols-r}) causes the following TOA residual,
%\begin{equation}\label{toaresin}
%   \begin{split}
%  R(t)=&-\frac{\Omega-q\hat q\cdot\hat n}{2\Omega^2}\sigma\varphi_0\left[\sin \Omega t-\sin(\Omega t-\Omega L-qL\hat q\cdot\hat n)\right]\\
%  &+\frac{e_{jk}\hat n^j\hat n^k}{2\omega(1+\hat q\cdot\hat n)}\left[\sin\omega t-\sin\omega(t-L-L\hat q\cdot\hat n)\right],
%  \end{split}
%\end{equation}
%for a pulsar located at a distance of $L$ and at the direction indicated by a unit vector $\hat n$, where $q=\sqrt{\Omega^2-m^2}$ and $\hat q$ is the propagation direction of the GW.
The TOA residuals of any pair of pulsars (named $a$ and $b$) are correlated.
The cross-correlation function is defined to be $C(\theta)=\langle \tilde{R}_a(t)\tilde{R}_b(t)\rangle$ where $\theta$ is the angular separation between $a$ and $b$,
and the brackets indicate the ensemble average over the stochastic background.
Figure~\ref{fig-cor} shows the normalized correlation function $\zeta(\theta)=C(\theta)/C(0)$
induced by the massless field $\tilde h_{\mu\nu}$ (the left panel) and the scalar field $\varphi$ (the right panel).
The curve on the left panel is actually the same as the one in GR. On the right panel, we calculated $\zeta(\theta)$
for the massless (labeled by Breathing) and the massive (3 different masses in units of $m_b$) cases.
Therefore, $\zeta(\theta)$ has rather different behaviors for $\tilde h_{\mu\nu}$ and $\varphi$.
In addition, $\zeta(\theta)$ induced by $\varphi$ is quite sensitive to small masses
with $m\lesssim m_b$, while for  larger masses, $\zeta(\theta)$ barely changes.
So this provides the possibility to determine the polarization content of GWs.
In Ref.~\cite{Lee2013}, Lee also calculated the cross-correlation functions and
his results (the right two panels in his Figure 1) are different from those on the right panel
in Figure~\ref{fig-cor}, because in his treatment,
the longitudinal and the transverse polarizations were assumed to be two independent modes.
In our approach, it is not permissible to calculate the cross-correlation function separately for the longitudinal and the transverse polarizations, as they are both excited by the same field $\varphi$ and the polarization state is a single mode.

\section{Conclusion}

We first studied the physical degrees of freedom in $f(R)$ gravity with two different approaches.
Both of them give the same result: three physical propagating degrees of freedom.
We then solved the linearized equations of motion for $f(R)$ gravity and obtained the plane wave solution.
The geodesic deviation equations are computed to reveal the polarizations,
and there are longitudinal and transverse breathing polarizations, in addition to the $+$ and $\times$ polarizations.
We also extended the work in Ref.~\cite{Liang:2017gwa} to Horndeski theory \cite{Hou:2017gwb}.
The analysis shows that the scalar field excites both the longitudinal and the transverse breathing polarizations if it is massive,
while it excites only the transverse breathing polarization if it is massless.
We find that $\Psi_2$ is zero in both $f(R)$ gravity and Horndeski theory, and that the longitudinal polarization exists even though $\Psi_2=0$.
The results show the failure of NP formalism \cite{Eardley1973} in classifying the polarizations of non-null GWs.
Since the six polarization states are completely determined by the electric part of the Riemann tensor $R_{itjt}$, we can still use the six
polarizations classified by the NP formalism as the base states. In terms of these polarization base states,
the polarization state caused by the massive
scalar field is a mix of the longitudinal and the breathing modes.
The interferometer responses functions were then computed and found out that it is difficult for interferometers to detect the longitudinal polarization.
We also predicted the cross-correlation functions. It implies the possibility of using pulsar timing arrays to detect the polarizations caused by the scalar field.

\textbf{Acknowledgements}:
We would like to thank Ke-Jia Lee for useful discussions.
This research was supported in part by the Major Program of the National Natural Science Foundation of China under Grant No. 11690021 and the National Natural Science Foundation of China under Grant No. 11475065.

%
% BibTeX or Biber users please use (the style is already called in the class, ensure that the "woc.bst" style is in your local directory)
% \bibliography{name or your bibliography database}

%\bibliographystyle{woc}
%\bibliography{../../References/references}

\begin{thebibliography}{58}

\bibitem{gw150914}
B.P. Abbott et~al. (LIGO Scientific Collaboration and Virgo Collaboration),
  Phys. Rev. Lett. \textbf{116}, 061102 (2016)

\bibitem{gw151226}
B.P. Abbott et~al. (LIGO Scientific Collaboration and Virgo Collaboration),
  Phys. Rev. Lett. \textbf{116}, 241103 (2016)

\bibitem{gw170104}
B.P. Abbott et~al. (LIGO Scientific Collaboration and Virgo Collaboration),
  Phys. Rev. Lett. \textbf{118}, 221101 (2017)

\bibitem{Newman1962}
E.~Newman, R.~Penrose, J. Math. Phys. \textbf{3}, 566 (1962)

\bibitem{Eardley1973}
D.M. Eardley, D.L. Lee, A.P. Lightman, Phys. Rev. D \textbf{8}, 3308 (1973)

\bibitem{Liang:2017gwa}
D.~Liang, Y.~Gong, S.~Hou, Y.~Liu, Phys. Rev. D \textbf{95}, 104034 (2017)

\bibitem{Hellings1983}
R.W. Hellings, G.S. Downs, Astrophys. J. \textbf{265}, L39 (1983)

\bibitem{Utiyama1962}
R.~Utiyama, B.S. DeWitt, J. Math. Phys. \textbf{3}, 608 (1962)

\bibitem{Stelle1977}
K.S. Stelle, Phys. Rev. D \textbf{16}, 953 (1977)

\bibitem{Buchdahl1970}
H.A. Buchdahl, Mon. Not. Roy. Astron. Soc. \textbf{150}, 1 (1970)

\bibitem{Adam1998}
G.R. Adam et~al., Astron. J. \textbf{116}, 1009 (1998)

\bibitem{Perlmutter1999}
S.~Perlmutter et~al., Astrophys. J. \textbf{517}, 565 (1999)

\bibitem{Horndeski1974}
G.W. Horndeski, Int. J. of Theor. Phys. \textbf{10}, 363 (1974)

\bibitem{Ostrogradsky1850}
M.~Ostrogradsky, Mem. Acad. St Petersbourg \textbf{6}, 385 (1850)

\bibitem{Hou:2017gwb}
S.~Hou, Y.~Gong, Y.~Liu (2017), \texttt{1704.01899}

\bibitem{Corda2007}
C.~Corda, JCAP \textbf{2007}, 009 (2007)

\bibitem{Corda2008}
C.~Corda, Int. J. Mod. Phys. A \textbf{23}, 1521 (2008)

\bibitem{Kausar2016}
H.R. Kausar, L.~Philippoz, P.~Jetzer, Phys. Rev. D \textbf{93}, 124071 (2016)

\bibitem{Myung2016}
Y.S. Myung, Adv. High Energy Phys. \textbf{2016}, 3901734 (2016)

\bibitem{Starobinsky1980}
A.A. Starobinsky, Phys. Lett. B \textbf{91}, 99 (1980)

\bibitem{Planck2016}
P.A.R. Ade et~al. (Planck Collaboration), Astron. Astrophys. \textbf{594}, A20
  (2016)

\bibitem{Vollick2003}
D.N. Vollick, Phys. Rev. D \textbf{68}, 063510 (2003)

\bibitem{Nojiri2003}
S.~Nojiri, S.D. Odintsov, Phys. Rev. D \textbf{68}, 123512 (2003)

\bibitem{Carroll2004}
S.M. Carroll, V.~Duvvuri, M.~Trodden, M.S. Turner, Phys. Rev. D \textbf{70},
  043528 (2004)

\bibitem{Flanagan2004}
E.E. Flanagan, Phys. Rev. Lett. \textbf{92}, 071101 (2004)

\bibitem{Chiba2013}
T.~Chiba, Phys. Lett. B \textbf{575}, 1 (2003)

\bibitem{Erickcek2006}
A.L. Erickcek, T.L. Smith, M.~Kamionkowski, Phys. Rev. D \textbf{74}, 121501
  (2006)

\bibitem{Hu2007}
W.~Hu, I.~Sawicki, Phys. Rev. D \textbf{76}, 064004 (2007)

\bibitem{Starobinsky2007}
A.A. Starobinsky, JETP Letters \textbf{86}, 157 (2007)

\bibitem{Cognola2008}
G.~Cognola, E.~Elizalde, S.~Nojiri, S.D. Odintsov, L.~Sebastiani, S.~Zerbini,
  Phys. Rev. D \textbf{77}, 046009 (2008)

\bibitem{Nojiri2008}
S.~Nojiri, S.D. Odintsov, Phys. Rev. D \textbf{77}, 026007 (2008)

\bibitem{Capozziello2009}
S.~Capozziello, M.~De~Laurentis, S.~Nojiri, S.D. Odintsov, Gen. Relativ.
  Gravit. \textbf{41}, 2313 (2009)

\bibitem{Myrzakulov2015}
R.~Myrzakulov, L.~Sebastiani, S.~Vagnozzi, Eur. Phys. J. C \textbf{75}, 444
  (2015)

\bibitem{Yi2016}
Z.~Yi, Y.~Gong, Phys. Rev. D \textbf{94}, 103527 (2016)

\bibitem{Hanlon1972}
J.~O'Hanlon, Phys. Rev. Lett. \textbf{29}, 137 (1972)

\bibitem{Teyssandier1983}
P.~Teyssandier, P.~Tourrenc, J. Math. Phys. \textbf{24}, 2793 (1983)

\bibitem{Capozziello2008}
S.~Capozziello, C.~Corda, M.F. De~Laurentis, Phys. Lett. B \textbf{669}, 255
  (2008)

\bibitem{Capozziello2010}
S.~Capozziello, R.~Cianci, M.~De~Laurentis, S.~Vignolo, Eur. Phys. J. C
  \textbf{70}, 341 (2010)

\bibitem{Goldhaber1974}
A.S. Goldhaber, M.M. Nieto, Phys. Rev. D \textbf{9}, 1119 (1974)

\bibitem{Berry:2011pb}
C.P.L. Berry, J.R. Gair, Phys. Rev. D \textbf{83}, 104022 (2011), [Erratum:
  Phys. Rev. D 85,089906 (2012)]

\bibitem{Ezawa1999}
E.~Yasuo, K.~Masahiro, K.~Masahiko, S.~Jiro, Y.~Tadashi, Class. Quantum Grav.
  \textbf{16}, 1127 (1999)

\bibitem{Ezawa2006}
Y.~Ezawa, H.~Iwasaki, Y.~Ohkuwa, S.~Watanabe, N.~Yamada, T.~Yano, Class.
  Quantum Grav. \textbf{23}, 3205 (2006)

\bibitem{Deruelle2009}
N.~Deruelle, Y.~Sendouda, A.~Youssef, Phys. Rev. D \textbf{80}, 084032 (2009)

\bibitem{Deruelle2010}
N.~Deruelle, M.~Sasaki, Y.~Sendouda, D.~Yamauchi, Prog. Theor. Phys.
  \textbf{123}, 169 (2010)

\bibitem{Sendouda2011}
Y.~Sendouda, N.~Deruelle, M.~Sasaki, D.~Yamauchi, Int. J. Mod. Phys.:
  Conference Series \textbf{01}, 297 (2011)

\bibitem{Olmo2011}
G.J. Olmo, H.~Sanchis-Alepuz, Phys. Rev. D \textbf{83}, 104036 (2011)

\bibitem{Ohkuwa2015}
Y.~Ohkuwa, Y.~Ezawa, Eur. Phys. J. Plus \textbf{130}, 77 (2015)

\bibitem{Arnowitt1962}
R.~Arnowitt, S.~Deser, C.W. Misner, \emph{Gravitation: An Introduction to
  Current Research} (John Wiley and Sons Ltd., New York, 1962), pp. 227--265

\bibitem{arnowitt_republication_2008}
R.~Arnowitt, S.~Deser, C.W. Misner, Gen. Relativ. Gravit. \textbf{40}, 1997
  (2008)

\bibitem{Kobayashi2011ginf}
T.~Kobayashi, M.~Yamaguchi, J.~Yokoyama, Prog. Theor. Phys. \textbf{126}, 511
  (2011)

\bibitem{Rakhmanov2005}
M.~Rakhmanov, Phys. Rev. D \textbf{71}, 084003 (2005)

\bibitem{Lee2008ptac}
K.J. Lee, F.A. Jenet, H.P. Richard, Astrophys. J. \textbf{685}, 1304 (2008)

\bibitem{Lee2010pta}
K.~Lee, F.A. Jenet, R.H. Price, N.~Wex, M.~Kramer, Astrophys. J. \textbf{722},
  1589 (2010)

\bibitem{Chamberlin2012}
S.J. Chamberlin, X.~Siemens, Phys. Rev. D \textbf{85}, 082001 (2012)

\bibitem{Lee2013}
K.J. Lee, Class. Quantum Grav. \textbf{30}, 224016 (2013)

\bibitem{Yunes2013lrr}
N.~Yunes, X.~Siemens, Living Rev. Relativity \textbf{16}, 9 (2013)

\bibitem{Gair2014}
J.~Gair, J.D. Romano, S.~Taylor, C.M.F. Mingarelli, Phys. Rev. D \textbf{90},
  082001 (2014)

\bibitem{Gair2015}
J.R. Gair, J.D. Romano, S.R. Taylor, Phys. Rev. D \textbf{92}, 102003 (2015)

\end{thebibliography}

\end{document}